# Histopathology for Mohs Micrographic Surgery with Photoacoustic Remote Sensing Microscopy

BENJAMIN R. ECCLESTONE,[1] KEVAN BELL,[1,2] SAAD ABBASI,[1] DEEPAK DINAKARAN,[2,3] MUBA TAHER,[4] JOHN R. MACKEY,[2,3] PARSIN HAJI REZA[1,*]

[1]PhotoMedicine Labs, Department of Systems Design Engineering, University of Waterloo, 200 University Ave W, Waterloo, Ontario, N2L 3G1, Canada
[2]illumiSonics, Inc., Department of Systems Design Engineering, University of Waterloo, 200 University Ave W, Waterloo, Ontario, N2L 3G1, Canada
[3] Department of Oncology, University of Alberta, 8440 112 St. NW, T6G 2R7, Edmonton, Alberta, Canada
[4] Division of Dermatology, Department of Medicine, University of Alberta, 116 St & 85 Ave, Edmonton, Alberta, T6G 2V1, Canada

*phajireza@uwaterloo.ca

**Abstract:** Mohs micrographic surgery (MMS) is a precise oncological technique where layers of tissue are resected and examined with intraoperative histopathology to minimize the removal of normal tissue while completely excising the cancer. To achieve intraoperative pathology, the tissue is frozen, sectioned and stained over a 20- to 60-minute period, then analyzed by the MMS surgeon. Surgery is continued one layer at a time until no cancerous cells remain, meaning MMS can take several hours to complete. Ideally, it would be desirable to circumvent or augment frozen sectioning methods and directly visualize subcellular morphology on the unprocessed excised tissues. Employing photoacoustic remote sensing (PARS™) microscopy, we present a non-contact label-free reflection-mode method of performing such visualizations in frozen sections of human skin. PARS leverages endogenous optical absorption contrast within cell nuclei to provide visualizations reminiscent of histochemical staining techniques. Presented here, is the first true one to one comparison between PARS microscopy and standard histopathological imaging in human tissues. We demonstrate the ability of PARS microscopy to provide large grossing scans (>1 cm$^2$, sufficient to visualize entire MMS sections) and regional scans with subcellular lateral resolution (300 nm).

## 1. Introduction

Mohs micrographic surgery (MMS) is the gold standard precision surgical technique for treating contiguous invading skin cancers in cosmetically and functionally important areas [1]. MMS excision of nonmelanoma skin cancers (NMSC) represents one of the most common procedures in the United States. Around 25% of the 3.5 million NMSC cases diagnosed each year are treated with this procedure [2,3]. In recent years NMSC incidence has risen dramatically, straining the global capacity to provide MMS [2]. For the two most common NMSCs, basal cell carcinomas (BCC) and squamous cell carcinomas (SCC), MMS achieves a five-year cure rate of nearly 99% [4,5]. For high risk nonmelanoma lesions, MMS achieves higher cure rates than wide local excision [6-8].

During MMS the surgeon repeatedly excises thin tissue layers which then undergo intraoperative histopathological analysis to identify regions of invasion at the margins. Each layer of tissue will be about ~5 mm thick and will aim to capture a 2 to 3 mm margin around the tumor [9]. These excised tissue samples will then undergo frozen histopathology. Standard frozen histology consists of embedding the sample into a cutting substrate, then cooling the sample and substrate to -25°C in a cryostat. Once frozen, the inner surface of the sample is sectioned via cryotome into 5-10-micron slices and placed onto a microscope slide. These slides are then dyed with histochemical stains to provide contrast for microscopic assessment. In contrast to other surgical techniques, in MMS the entire deep and peripheral margin of the excised tissue undergoes pathological analysis. By assessing the entire surgical margin in this manner, the surgeon is able to identify specific invasive regions of malignant cells, which are then targeted during the next excision. This process of layer by layer excision with interim histopathological analysis is repeated until the entire invasive tumor has been removed [9].

The use of intraoperative frozen histology means operating times for MMS can exceed several hours, depending on the number of excisions required [10]. Most of this time is consumed by tissue processing as described, which requires 20 to 60 minutes per individual tissue layer, while the excision and pathologic assessment are relatively rapid [9]. Consequently, the rate limiting step in the MMS process is generating the tissue suitable for transmission light microscopic interpretation. The MMS process could be accelerated greatly if tissue processing could be circumvented, and unstained tissues visualized directly. Ideally, images so obtained would be visually similar to standard histochemical staining images. Furthermore, this would also preserve sample integrity permitting re-examination and additional immunochemistry analysis.

Direct histological imaging of unstained tissue sections presents unique challenges compared to pathological analysis of stained frozen preparations. Nonetheless, a variety of imaging methods have demonstrated histology-like imaging in MMS excisions. Prominent examples include microscopy with ultraviolet surface excitation (MUSE) [11], multiphoton fluorescence microscopy (MPM) [12,13], Raman spectroscopy [14-16], photoacoustic microscopy (PAM) [17,18], and optical coherence tomography (OCT) [19-21], each of which has been explored for intraoperative histology during MMS. While MUSE and MPM have shown promising histology-like images, they require exogenous dyes for contrast. Staining tissues prior to imaging reintroduces many of the tissue preparation issues experienced by frozen histology as staining can be resource intensive and introduces potential for variability.

Only Raman spectroscopy, PAM and OCT have demonstrated label-free histology-like imaging of MMS specimens [14-21]. Unfortunately, recent works on MMS histology with PAM [17,18] and Raman spectroscopy [14-16] feature inferior resolutions compared to conventional optical microscopy. Though PAM [22, 23] and Raman spectroscopy [24] may provide subcellular resolution this has not been applied in MMS. Hence, the current embodiments [14-18] are inadequate for precisely locating small and subtle regions of malignant cells. Of the presented methods, OCT is the only method which provides subcellular resolution and label free contrast. However, the optical scattering contrast in OCT does not provide the specificity necessary to match current pathology standards [19-21]. To provide visualizations reminiscent of current H&E staining, OCT systems must use external image processing techniques. Therefore, there remains a pressing need for an accurate interoperative label-free histopathological microscopy technique capable of imaging large areas of tissue while also providing subcellular resolution to expedite MMS procedures.

Photoacoustic Remote Sensing (PARS™) microscopy has recently emerged as an all-optical non-contact label-free reflection-mode imaging modality [25-27]. Like other photoacoustic techniques, PARS captures endogenous optical absorption contrast visualizing a wide range of biological chromophores including hemoglobin, lipids, and DNA. A pulsed excitation laser is used to deposit optical energy into the sample. As the target chromophore absorbs the excitation pulse, it undergoes thermo-elastic expansion proportional to the absorbed excitation energy. The expansion induces nanosecond scale modulations in the local refractive index of the sample. In PARS, this effect is observed as back-reflected intensity variations of a second co-focused continuous-wave detection laser. In this way, PARS microscopy visualizes endogenous absorption contrast in an all-optical label-free reflection-mode architecture [25-27]. Previously, our group has shown promising histology-like imaging capabilities by utilizing ultraviolet (UV) excitation to primarily target the absorption contrast of DNA [26,27]. Accentuating this contrast generates visualizations reminiscent of immunohistochemical staining of cell nuclei. In recent works, PARS histological imaging has been applied in thick tissue samples (>2 mm) including freshly resected tissues and formalin fixed paraffin embedded tissue preparations [26,27,28].

In this work, we present a PARS system for rapid label-free histological imaging of unprocessed MMS sections. By leveraging recent technical improvements of the PARS system, we have expanded the histological imaging capabilities to match the scanning area, resolution and imaging speed required for MMS. We show the first non-contact photoacoustic microscopy of Mohs excisions. Unstained MMS excisions are imaged with the PARS microscope, then stained and imaged with a brightfield microscope. Using this strategy, we provide the first true one to one comparison between PARS microscopy and normal histopathological imaging. Wide field of view grossing scans capture entire MMS sections (>1 cm$^2$ area) with sufficient resolution to recover subcellular diagnostic characteristics. Concurrently, smaller high-resolution images give close-ups of clinically relevant regions. These small fields provide ~300 nm optical resolution enabling morphological assessment of single nuclei. Thus, the proposed PARS system provides both the grossing scan capabilities, and high spatial resolution required to assess tissues during MMS. Compared to frozen histology, the presented PARS microscope without optimized scanning hardware can image an entire MMS excision in under 12 minutes, 60% of the time to prepare a slide for brightfield histological assessment. Applied in a clinical setting, this device may circumvent the need for histopathological processing of tissue. Ideally, the PARS system could be implemented into standard histopathological workflows without affecting current techniques. Thus, PARS is well positioned to supplement existing intraoperative tissue analysis techniques, potentially reducing the time for each MMS operative cycle, streamlining the MMS process, thereby increasing MMS capacity.

## 2. Methods

2.1 *Imaging System:*

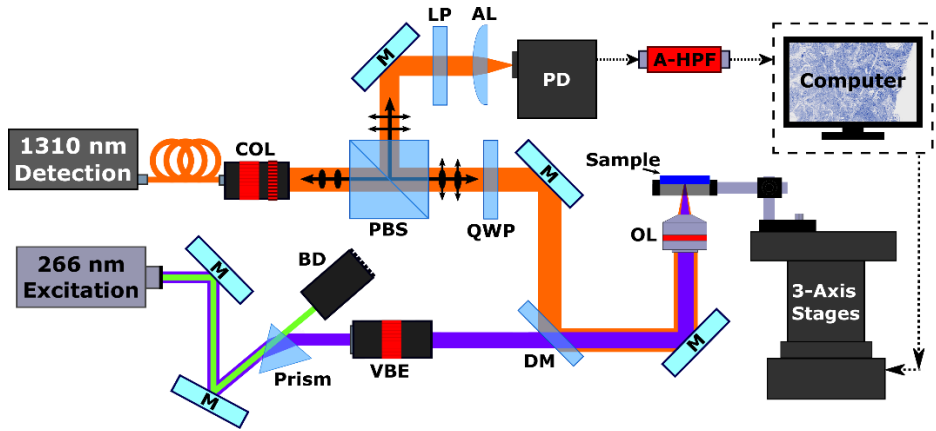

Fig. 1. Simplified Schematic of the PARS system. Component labels are defined as follows: collimator (COL), polarizing beam splitter (PBS), quarter wave plate (QWP), dichroic mirror (DM), variable beam expander (VBE), beam dump (BD), objective lens (OL), long pass filter (LP), aspheric focal lens (AL), photodiode (PD), analogue high-pass filter (A-HPF), mirrors (M).

The proposed imaging system is shown in Figure 1. A 400 ps pulsed laser (WEDGE XF, Bright Solutions) was selected to provide 266 nm UV excitation. Residual 532 nm output is removed from the excitation beam with a CAF2 prism (PS862, Thorlabs). Following separation, the UV beam is expanded (BE05-266, Thorlabs) and combined with the detection beam via dichroic mirror (HBSY234, Thorlabs). Detection of PARS signals is performed with a 1310 nm continuous wave super-luminescent diode (S5FC1018P, Thorlabs). Collimated, horizontally polarized detection light passes through a polarizing beam splitter (PBS254, Thorlabs) and quarter wave plate (WPQ10M-1310, Thorlabs) into the imaging system. Both excitation and detection are then co-focused onto the sample with a 0.5 NA reflective objective (LMM-15X-UVV, Thorlabs). The back reflected detection beam from the sample which encodes the PARS modulations, is returned along detection pathway. Passing through the quarter wave plate for a second time, the detection beam becomes horizontally polarized. The horizontally polarized light is then directed towards the photodetector by the polarizing beam splitter where it is filtered (FELH1000, Thorlabs) before being focused onto the photodiode (PDB425C-AC, Thorlabs).

Images were collected by mechanically scanning samples over the fixed imaging head in a continued raster pattern. While scanning, the excitation laser pulses continually capturing evenly spaced PARS interrogation points. The lateral spacing between PARS collection points is then tuned by adjusting the stage speed and excitation laser repetition rate. Depending on the desired resolution, the lateral spacing ranged from 0.1 to 5 µm. Each time the excitation laser is pulsed, a short segment of photodetector signal is recorded capturing the PARS modulations. To ensure accurate recovery of the PARS interrogation, around 250 samples of the photodiode signal are captured with a 14-bit digitizer (RZE-004-300, Gage Applied). This time domain data is streamed via PCI-E channel from the digitizer to the computer memory. Here, an algorithm was applied to extract the characteristic amplitude of each PARS signal in real time.

## 2.2 Image Reconstruction and Processing

At each interrogation a positional signal from the stage is collected along with the characteristic PARS amplitude. The positional signal is then used to remove data with irregular spatial sampling characteristics (i.e. data collected while the stage was accelerating). Once the sections with irregular spatial sampling are removed, the remaining data forms a perfect cartesian grid of PARS signals. This grid data is essentially a raw image ready for further processing [26,27]. Once a raw frame has been formed, some standard processing steps are performed to generate a PARS image. First, to reduce measurement noise, the data is gaussian filtered. Then, to enable consistent processing between images, the raw data is normalized. The absorption data is then rescaled logarithmically, enabling visualization of absorption contrast across different orders of magnitude. Logarithmic rescaling is performed by converting the normalized linear PARS signal into decibels. Once converted to log space, the data scaling is adjusted based on the histogram distribution to reduce background noise around the tissue. This is the complete processing for the small high-resolution frames. Large field of view frames undergo one further refinement. For large frames, flat field correction is applied to reduce contrast artifacts. Flat field correction is applied using a gaussian smoothing approach. Finalized images, are then exported as image files which can be easily viewed and analyzed.

2.3 *Sample Preparation:*

In this study, a variety of MMS excisions with BCC were selected for imaging. Frozen sections of tissue specimens with BCC were obtained from Mohs surgeries. These specimens underwent standard Mohs sample preparation as follows. The Mohs excisions are embedded within a cutting substrate and placed into a cryostat where they are cooled to approximately -25°C, over a 1 to 10-minute period (depending on sample shape and composition). The frozen samples were then sectioned via cryotome into 5-10-micron slices and placed onto a microscope slide. The slide was then dried and fixed at 55°C for 1 minute. The unstained tissue slices were then imaged at room temperature with the PARS microscope. Following PARS imaging, the slides were returned to the clinicians to undergo the remaining standard processing. Hematoxylin and eosin (H&E) staining was performed, then the slides were covered with mounting media and a cover slip. Following processing the now stained slices were imaged with a standard brightfield microscope (Zeiss Axioscope 2 with Zeiss Axiocam HR). The tissue samples were collected under protocols approved by the Research Ethics Board of Alberta (Protocol ID: HREBA.CC-18-0277) and the University of Waterloo Health Research Ethics Committee (Humans: #40275). The ethics committees waived the requirement for patient consent as the selected samples were excess tissues no-longer required for diagnostic purposes, and no patient identifiers were provided to the researchers. All experiments were performed in accordance with the requirements of the Government of Canada Panel on Research Ethics Guidelines.

3. **Results and Discussion**

Prior to imaging human tissue samples, the microscope was characterized using gold nanoparticles and polystyrene microspheres. Gold nanoparticles were used to measure the lateral optical resolution of the system, while the polystyrene microspheres were used to investigate the relationship between resolution and spatial sampling. The systems axial resolution was not characterized for this study. Applied in thick tissues, optical sectioning can be used to recover signals at different depths within the sample [26,27,28]. However, the thin samples used in this study, do not provide a depth resolvable phantom since PARS signal is generated over the entire thickness with each excitation. The lateral optical resolution was determined from the full width half maximum (FWHM), of imaging the 200 nm nanoparticles. To fully capture the nanoparticles, the minimum accurate lateral step sizes of 25 nm and 50 nm for the x and y lateral step respectively were used (Figure 2a). Based on the average FWHM of 50 gold nanoparticles, the resolution was determined to be ~300 nm (Figure 2a). While the optical resolution provides an ideal metric, in point scanning microscopy the spatial sampling in addition to the optical resolution will affect the final image resolution.

To determine the effects of the spatial sampling interval on image formation, polystyrene microspheres were imaged with varying lateral step sizes. The 0.95 $\mu m$ microspheres were selected as their size is representative of cell nuclei. While imaging these samples, the spatial sampling rate was varied from 25 nm to 250 nm. The effects of varying the lateral step size can be seen in Figure 2b. As the sampling rate is decreased the resolving power is correspondingly decreased. The spatial resolution is ultimately determined by a combination of the optical resolution and the spatial sampling (pixel resolution). If the optical resolution is finer than the pixel size, the pixel size determines the image resolution. Alternatively, if the optical resolution is larger than the pixel size, the optical resolution will be dominant. In the large grossing scans with 4 $\mu m$ pixel size, the spatial sampling rate determines the resolution. In the smaller high-fidelity frames with 250 $nm$ pixel size ($< 300$ nm optical resolution), the optical resolution is dominant. For this reason, 250 $nm$ spatial sampling rate was selected for the high-fidelity imaging. Using these steps, micron scale structures can be resolved with near optical resolution while maintaining lower scanning times and smaller data volumes.

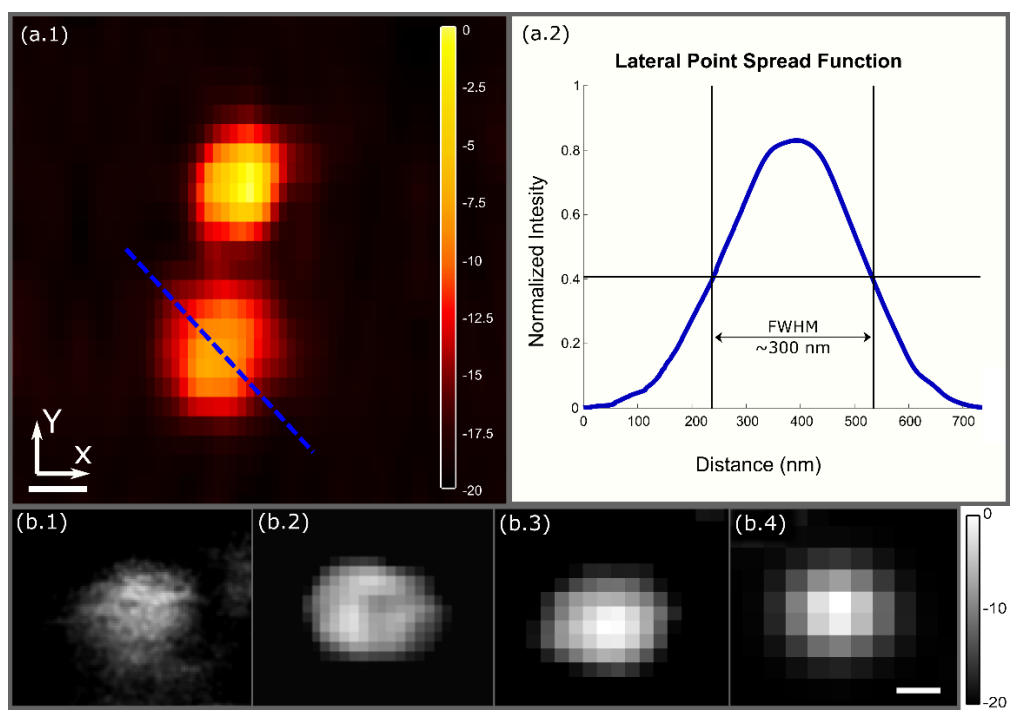

Fig. 2. Resolution characterization of the PARS system. a) (1) Image of 200 nm diameter gold nanoparticles acquired with the PARS system using a 25 nm lateral step size in the x, and a 50 nm lateral step size in the y. (Scale bar: 100 nm) (Dynamic range presented in decibels) (2) Lateral point spread function of the PARS system from gold nanoparticles, averaged across 50 nanoparticles. (FWHM resolution: ~300 nm) b) PARS image of a 0.95 μm polystyrene bead used to test the spatial sampling rates. (1) acquired using a 25 nm lateral step size. (2) acquired using a 75 nm lateral step size. (3) acquired using a 175 nm lateral step size. (4) acquired using a 250 nm lateral step size. (Scale bar: 500 nm) (Dynamic range presented in decibels).

Moving towards clinical applications, it would be ideal to incorporate the PARS system into the current tissue processing scheme. PARS could potentially be used to image tissues directly after excision, prior to histological analysis. Thus, nuclear morphology could be recovered from bulk tissues immediately during surgery, while still allowing further analysis and immunochemistry. To be implemented in this fashion, PARS should not interfere with current processing techniques. Imaging artifacts such as modification or degradation of tissues must be avoided. Therefore, the PARS optical system was refined to maximize imaging sensitivity and minimize the required excitation energy. Two avenues of optimization were pursued in this refinement, efficiency of photoacoustic generation and efficacy of PARS modulation recovery. To increase the localized photoacoustic pressure without increasing pulse energies, a 0.5 NA objective lens was exchanged for the previously utilized 0.3 NA lens [26,27]. Concurrently, the beam paths were condensed to the shortest viable path lengths. Reducing path lengths decreases vibration and thermal sensitivity. Shortening the beam paths also reduces the relative lateral displacement of the beam at the objective per microradian rotation in each alignment mirror. This allows for more precise manual positioning and co-alignment of the detection and excitation foci.

Additionally, further refinements were made to the detection pathway. Since PARS measures photoacoustic pressures as a modulation in the back reflected intensity of the detection beam, the detection sensitivity inherently depends on the efficiency of back reflection to the photodiode. Each percentage increase in back reflection efficiency is accompanied by a corresponding increase in the imaging sensitivity. Therefore, the optical system was refined to optimize the return of detection beam from the sample to the photodiode. There were two goals of the detection path refinements. First, to remove all non-essential optical components. Second, to reduce the detection power losses at each essential optical element. To this end, the galvo-mirrors and a dichroic mirror used in previous PARS embodiments [26, 27] were removed, resulting in an >15% increase in returned power through the detection path. Cumulatively, these refinements have enhanced contrast and reduced the excitation energy required for imaging. We have accentuated subtle contrast within tissue specimens, while reducing the risk of damaging tissues. This corresponded to a decrease in excitation pulse energy to about 750 pJ, from other reports of ~5 nJ [29].

Applying this system to unstained tissue samples, results of large grossing scans are shown in Figure 3. Figure 3a and 3b feature large field of view scans of a human MMS excision with BCC (13 mm and 10 mm side lengths respectively). The larger 13 mm scan was collected in ~12 minutes, while the smaller 10 mm scan was collected in ~8 minutes. In each case, dense regions of tumor tissues are observed within the excised tissues. Moreover, the bulk tissue morphology and the surgical margins can be assessed readily. A region of abnormal tissue in the excision sample, consisting of hypercellularity and architectural distortion, is seen extending and infiltrating into normal skin tissue (red outline, Figure 3a). Further resection layer shows only normal skin tissue (Figure 3b) with hair follicles being visible (green circles).

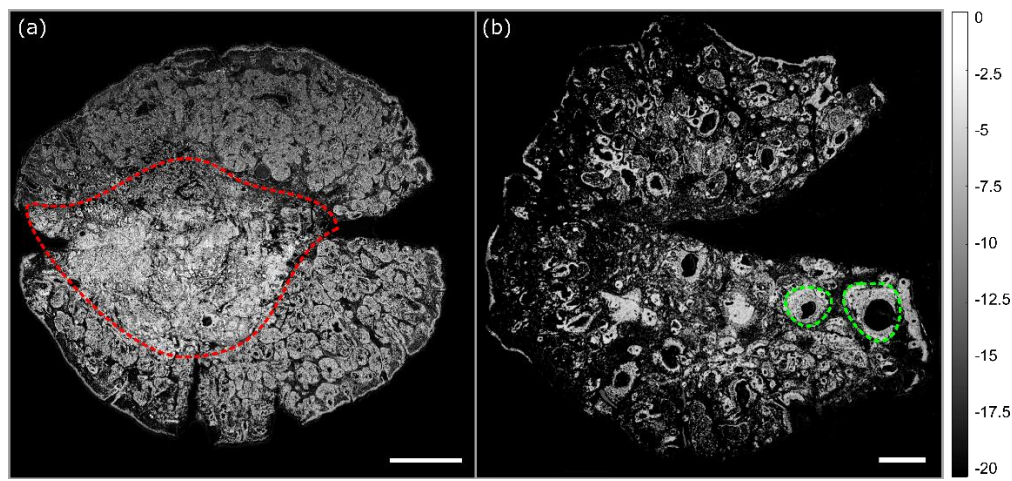

Fig. 3. Wide field of view PARS images of entire Mohs excisions. a) 13 mm by 13 mm PARS image of human skin tissue with basal cell carcinoma (red outline) shown by the increased cellularity in the middle (deep) margin causing invasion and architectural distortion of the normal skin (scale Bar: 2 mm). b) 10 mm by 10 mm PARS image of human skin tissue, where two examples of hair follicles are circled in green (scale Bar: 1 mm). Both a & b feature a 4 µm lateral step size. The notch in the tissue signifies the superior margin and aids in orientation. (Dynamic range presented in decibels).

Assessing the entire excision specimen is crucial to the high success of MMS compared to other techniques. Observing the entire surgical margin means areas of contiguous invasion can be identified for excision. In this specimen, the initial resection (Figure 3a) showed areas suspicious for tumor invasion at the deep and inferior margin, which can be seen due to the increased cellularity, morphologically different cell structure, and region of architectural distortion surrounding the tumor. By necessity, tissue sections exceeding 1.0 $cm^2$ in area, such as the scans presented in Figure 3, must be imaged with subcellular resolution. Moreover, to improve on frozen sectioning, imaging must be performed in under 20 minutes. Ideally the imaging time should be further reduced, to under 10 minutes. Several technical improvements have been made over previous reports of PARS devices to match the scanning area, resolution and imaging speed required for MMS. These have included improvements to scanning accuracy, imaging speed, image reconstruction techniques, and data management. By the nature of the point scanning mechanism, these improvements have enhanced both the grossing scans and small field scans. At each scale, (large or small scan window) the number of interrogation points may remain the same, however, the excitation rate and/or mechanical scanning speed may be scaled up or down to provide the desired lateral step size and resolution.

Increasing the scan area is usually accompanied by a corresponding reduction in resolution or an increase in acquisition time. Fortunately, in PARS, the impacts on resolution and imaging speed may be mostly mitigated by selecting a sufficiently high excitation laser repetition rate. However, increasing the repetition rate while maintaining the same spatial sampling rate also necessitates an increase in the movement speed of the scanning stages. In the proposed system, these stages are limited to a maximum velocity of 200 mm/s. Thus, for the grossing scans displayed in Figure 3 a and b, maintaining 4 µm spatial sampling requires a repetition rate of 50 kHz. Operating at the maximum mechanical velocity and 50 kHz excitation, a 16-million-point scan similar to Figure 4, or Figure 5, could be captured in around 12 minutes. Moving forwards, the imaging time will be drastically reduced by exchanging the mechanical stages and excitation laser. Utilizing a commercially available 600 mm/s mechanical stage in conjunction with a 1 MHz excitation would reduce the imaging time from 12, to under 2 minutes. However, another issue arises as capturing images such as Figure 3 through Figure 5 (16-million-points/image) would potentially result in nearly 30 GB of data for a single capture. To circumvent this issue, an algorithm was applied to extract the characteristic amplitude of each PARS signal in real time. Thus, the memory requirement is reduced by around 256 times. As a result, the number of PARS signals which can be reasonably recovered in a single scan is increased by the same factor, enabling practical acquisition of the presented larger and/or higher resolution scans.

Compared to the wide field images in Figure 3, the tighter spatial sampling and smaller field of view in Figure 4 reveal more intricate tissue morphology. By reducing the lateral spatial sampling rate to 250 nm, we maintain our optically limited resolution of ~300 nm. Presented in Figure 4 is a series of high-resolution PARS images of a tissue sample with BCC (bottom: a.2, b.2, c.2). Brightfield images of the same tissue following preparation with H&E staining are shown across the top row (a.1, b.1, c.1) of Figure 4. The H&E staining in Figure 4 is perhaps the most common immunohistochemical stain set and is regularly used in MMS for assessing NMSC. Within the images, the margin of invasive disease is identified due to high cellularity (lower right of red boundary) and corresponding nuclear content on PARS imaging, which corresponds to the light microscopy images. The nuclear atypia, high nuclear-to-cytoplasm ratio, and disorganized cellular organization are indicative of the presence of cancerous tissue (c.2) and compare favorably to the findings under light microscopy (c.1). We emphasize, that the H&E images (Figure 4a.1, b.1, c.1) are of the exact same tissue imaged with the PARS system (Figure 4a.2, b.2, c.2). By imaging the unstained tissues with the PARS system, then staining the tissues with H&E and imaging with a brightfield microscope, we are able to assess both the accuracy of the PARS and the effects PARS might have on normal pathological analysis. This is the first time such a comparison has been done in human tissues. Observing the brightfield images (Figure 4a.1, b.1, c.1), there is no visible degradation in tissue or nuclear structures following PARS imaging. Therefore, it shows that the PARS imaging has not affected the ability to perform staining and pathological analysis on the tissues. This implies the PARS system could be used to augment the current histopathology workflow with little to no impact.

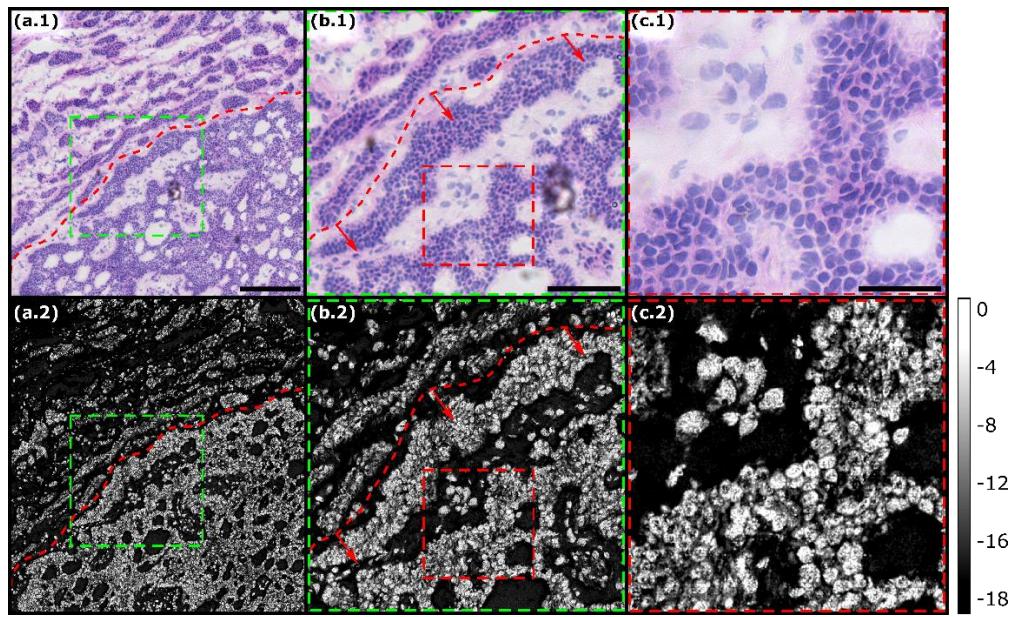

Fig.4 True one to one comparison of PARS and bright-field images of hematoxylin and eosin (H&E) stained human skin tissue with basal cell carcinoma (BCC). a) (1) 5x bright field image of tissue with BCC demonstrating the border of invasive cancer (bottom of red border) versus normal tissue (top of red border). (2) PARS image of the same unstained sample with BCC, the same red border denotes the cancer boundary. Scale Bar: 200 µm b) (1) 20x bright field image demonstrating the same cancer margin as in (a.1, a.2). (2) Enlarged section (green box) of the PARS image (a.2) compares the disorganized cellular architecture seen in the light microscopy (b.1) and PARS images (the same red border separates cancerous and normal tissues). Scale Bar: 100 µm c) (1) 20x bright field of tissue with clearly identifiable atypical nuclear morphology, size and distribution. (2) Enlarged section (red box) of the PARS image (b.2), providing a near perfect match to the brightfield histology image (c.1). Scale Bar: 40 µm. (Dynamic range presented in decibels).

Presented in Figure 5a is a series of 4 smaller 1 mm$^2$ images co-registered to form a 4 mm x 1 mm image with resolution equivalent to Figure 4. Each individual frame is 28 megapixels, 4000 by 7000-point image. Shown in this image is the transition (demarcated by the dashed red line) from cancerous tissue to normal tissue at the tumor margin, which is clearly visible by the different tissue architecture detectable by PARS imaging. The cancer cell's atypical cell and nuclear features are seen on higher magnification images Figure 5a to c. The ability to identify this tumor margin will guide the MMS surgeon in resecting the next layer of tissue.

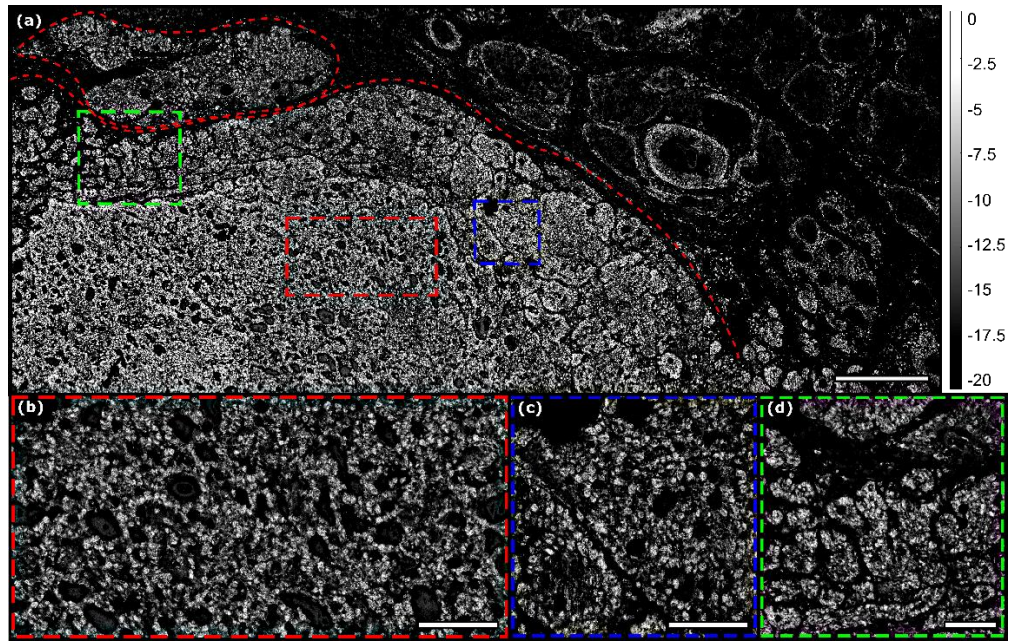

Fig 5. Large area high resolution PARS image of human skin tissue with basal cell carcinoma. (a) A series of four PARS images stitched together (112-megapixel image, 16000 by 7000-point scan, 250 nm step size) with hypercellularity and nuclear content. Evident disorganized cellular architecture denoting cancerous tissue is enclosed in the red border on the left side with normal tissue in the top and right. Scale Bar: 400 μm (b) Cropped and enlarged section (red box) of the PARS image shown in (a) Scale Bar: 100 μm (c) Cropped and enlarged section (blue box) of the PARS image shown in (a) Scale Bar: 100 μm (d) Cropped and enlarged section (green box) of the PARS image shown in (a) Scale Bar: 100 μm. (Dynamic range presented in decibels).

To capture absorption contrast, PARS leverages the photoacoustic effect. In PARS, a pulsed excitation laser is co-focused onto a sample with a continuous wave detection laser. The pulsed excitation laser is then used to deposit optical energy into the sample. As the target absorbs the excitation energy it undergoes thermoelastic expansion, proportional to the absorbed energy. The thermoelastic expansion causes corresponding modulations in the back-reflection of the detection beam via the elasto-optic effect. The modulations in the back-reflected detection beam are then proportional to the optical absorption [25-27]. In this application, the 266 nm excitation is selected mainly to target the optical absorption of DNA. This is an appropriate selection since DNAs UV absorption is orders of magnitude higher than most biological tissues. However, most common biomolecules have non-zero optical absorption of UV. This means PARS contrast could be captured from a wide variety of biomolecules if the detection were sensitive enough to capture the signals. With the refinements to improve detection sensitivity in this work, we are able to recover absorption contrast from a variety of chromophores in addition to DNA. However, since the PARS signal is directly proportional to the optical absorption, the signals recovered from DNA are orders of magnitude higher than signals recovered from other biomolecules. Therefore, if the raw data is presented with linear scaling only DNA is visible. In order to target extranuclear contrast, the raw data is transformed logarithmically during the image formation process.

Logarithmic scaling enables non-nuclear chromophores with absorption across many orders of magnitude to be visualized. With this method the subtle contrast attributed to biomolecules, such as cytochrome, hemoglobin and collagen can be recovered from the absorption data. Such extranuclear details are observed in Figure 5. While these system refinements and processing techniques have enhanced non-nuclear contrast, the extranuclear chromophores cannot be individually identified as their UV absorptions are relatively similar. However, recovering bulk tissue contrast from extranuclear structures still imparts a visualization benefit, as bulk tissue morphology may be captured in addition to nuclear contrast with a single wavelength. Moving forward, additional chromophore specific excitation wavelengths will be explored. Rather than using UV excitation, the infra-red absorption of DNA may be targeted to provide similar histology-like contrast while improving in-situ compatibility. Additionally, further excitation wavelengths will be added providing selective contrast for biomolecules such as lipids and collagen.

In addition to adding more chromophore specificity, future works will focus on increasing imaging speed. Considering the current system, capturing a single grossing or small region scan requires around 12 minutes and provides a 16-megapixel image. This is approximately 60% the time required for frozen histology preparation. However, to recover wider swaths such as Figure 5a, the scanning time increases linearly with 4 frames requiring 48 minutes. While still comparable to frozen sectioning, the PARS imaging time exceeds that of frozen pathology depending on how many scans are required. Moving forward, the imaging speed may be improved by increasing the point capture/excitation rate. Previous works have reported excitation rates of nearly 1 MHz, which if employed, would reduce imaging time to less than 30 seconds per scan [30]. Concurrently, to avoid mechanical limitations, more efficient scanning methods will be implemented. Future work will focus on incorporating these improvements aiming to image entire MMS excisions in one capture, with the same area as the presented grossing scans (Figure 3), and the same spatial sampling as the presented high-resolution captures (Figure 4 and Figure 5).

## 4. Conclusion

The presented results demonstrate the first visualization of nuclear morphology in human tissue samples exhibiting BCC, using a non-contact photoacoustic microscopy technique. By imaging unstained tissues, then staining and analyzing the same tissues we provide the first true one to one comparison between PARS microscopy and brightfield histopathological imaging. Shown here, subcellular structures are recovered from entire MMS sections, enabling full tissue margin analysis. Concurrently, small regions are captured providing close-ups of clinically relevant features such as nuclear organization, density and morphology. The one to one results presented to show the efficacy of the optimized PARS system, but do not presented here are provide a thorough clinical comparison. Moving forwards, the system and method developed here will soon be used to conduct a randomized controlled trial with clinicians to fully explore the pathological accuracy.

Notably, since PARS emulates the contrast provided by common histochemical stains, suggesting that if adopted, there may be little requirement to retrain pathologists to interpret a new image type. Ideally, the PARS system may facilitate recovery of diagnostic details more rapidly than conventional techniques by eliminating tissue processing steps. Moving forwards, the non-contact label-free reflection-mode PARS microscope presented here, could potentially be applied directly to unprocessed MMS excisions. As shown, PARS does not degrade or modify tissues samples in any way, causing no detriment to histopathological processing. This suggests bulk tissues could be imaged immediately after excision, while still being preserved in their entirety enabling re-examination, further histopathological processing and immunochemistry. Thus, this device may circumvent the need for frozen pathology, or could act as an adjunct to the current processing stream. Overall, the PARS modality is well suited to intraoperative guidance of MMS and could reduce MMS cycle time, increase patient flow, and free up histopathology staff to perform other tasks. Moving forwards, the performance of this system will be examined in freshly resected bulk MMS excisions.

## Funding

## Acknowledgments

N/A

### *Disclosures*

KB: illumiSonics Inc. (F, I, E, P); DD: illumiSonics Inc. (I); JRM: illumiSonics Inc. (I); PHR: illumiSonics Inc. (F, I, P, S).

11. T. Yoshitake, M.G. Giacomelli, L.M. Quintana, H. Vardeh, L.C. Cahill, B.E. Faulkner-Jones, J.L. Connolly, D. Do and J.G. Fujimoto, "Rapid histopathological imaging of skin and breast cancer surgical specimens using immersion microscopy with ultraviolet surface excitation," Sci. Rep. **8,** 4476 (2018).

12. C. Longo, M. Ragazzi, M. Rajadhyaksha, K. Nehal, A. Bennassar, G. Pellacani and J.M. Guilera, "In Vivo and Ex Vivo Confocal Microscopy for Dermatologic and Mohs Surgeons," Dermatol. Clin. **34**(4), 497-504 (2016).

13. J. Paoli, M. Smedh, A. Wennberg and M.B. Ericson, "Multiphoton Laser Scanning Microscopy on Non-Melanoma Skin Cancer: Morphologic Features for Future Non-Invasive Diagnostics," J. Invest. Dermatol. **128**(5), 1248-1255 (2008).

14. M. Larraona-Puy, A. Ghita, A. Zoladek, W. Perkins, S. Varma, I.H. Leach, A.A. Koloydenko, H. Williams and I. Notingher, "Discrimination between basal cell carcinoma and hair follicles in skin tissue sections by Raman micro-spectroscopy," J. Mol. Struct. **993**(3), 57-61 (2011).

15. K. Kong, C.J. Rowlands, S. Varma, W. Perkins, I.H. Leach, A.A. Koloydenko, A. Pitiot, H.C. Williams, and I. Notingher, "Increasing the speed of tumour diagnosis during surgery with selective scanning Raman microscopy," J. Mol. Struct. **1073**, 58-65 (2014).

16. C.A. Lieber, S.K. Majumder, D.L. Ellis, D.D. Billheimer and A. Mahadevan-Jansen, "In vivo nonmelanoma skin cancer diagnosis using Raman microspectroscopy," Lasers. Surg. Med. **40**(7), 461-467 (2008).

17. U. Dahlstrand, R. Sheikh, A. Merdasa, R. Chakari, B. Persson, M. Cinthio, T. Erlov, B. Gesslein, and M. Malmsjo, "Photoacoustic imaging for three-dimensional visualization and delineation of basal cell carcinoma in patients," Photoacoustics. **18,** 100187 (2020).

18. A.B.E. Attia, S.Y. Chuah, D. Razansky, C. Jun Hui Ho, P. Malempati, U.S. Dinish, R. Bi, C. Yaw Fu, S.J. Ford, J. Siong See Lee, M.W. Ping Tan, M. Olivo, and S. Tien Guan Thng, "Noninvasive real-time characterization of non-melanoma skin cancers with handheld optoacoustic probes," Photoacoustics. **7**, 20-26 (2017).

19. S.A. Alawi, M. Kuck, C. Wahrlich, S. Batz, G. McKenzie, J.W. Fluhr, J. Lademann and M. Ulrich, "Optical coherence tomography for presurgical margin assessment of non-melanoma skin cancer – a practical approach," Exp. Dermatol. **22**(8), 547-551 (2003).

20. D. Rashed, D. Shah, A. Freeman, R.J. Cook, C. Hopper and C.M. Perrett, "Rapid ex vivo examination of Mohs specimens using optical coherence tomography," Photodiagnosis. Photodyn. Ther. **19**, 243-248 (2017).

21. C.S. Chan, T.E. Rohrer. "Optical Coherence Tomography and Its Role in Mohs Micrographic Surgery: A Case Report," Case. Rep. Dermatol. **4**(3), 269-274 (2012).

22. A. Danielli, K.L. Maslov, A. Garcia-Uribe, A.M. Winkler, C. Li, L. Wang, Y. Cheng, G.W. Dorn II, L.V. Wong, "Label-free photoacoustic nanoscopy" J. Biomed. Opt. **19**, 086006 (2014).

23. E.M. Strohm, M.J. Moore, M.C. Kolios, "High resolution ultrasound and photoacoustic imaging of single cells," Photoacoustics. **4**, 36 (2016).

24. Y. Bi, C. Yang, Y. Chen, S. Yan, G. Yang, Y. Wu, G. Zhang, P. Wang, "Near-resonance enhanced label-free stimulated Raman scattering microscopy with spatial resolution near 130 nm," Light Sci. Appl. **7**, 81 (2018).

25. P. Haji Reza, W. Shi, K. Bell, P. Paproski, and R.J. Zemp, "Non-interferometric photoacoustic remote sensing microscopy," Light. Sci. Appl., **6**, e16278 (2017).

26. S. Abbasi, M. Le, B. Sonier, K. Bell, D. Dinakaran, G. Bigras, J.R. Mackey and P. Haji Reza, "Chromophore selective multi-wavelength photoacoustic remote sensing of unstained human tissues," Biomed. Opt. Express. **10**(11), 5461-5469 (2019).
13

**Article Thumbnail:**

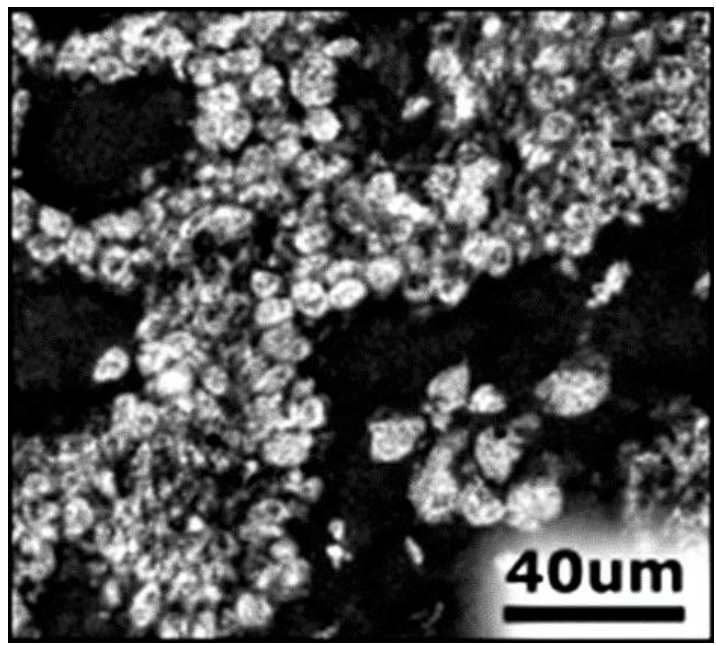